\documentclass{article}
\pdfoutput=1
\usepackage{graphicx}  
\usepackage{amsmath}   
\usepackage{amssymb}   
\usepackage{bm} 
\usepackage{dcolumn}
\usepackage{color}
\usepackage{mathrsfs}
\usepackage{amsfonts}
\usepackage{varioref}
\usepackage[utf8]{inputenc}
\usepackage{amsmath}
\usepackage{amsfonts}
\usepackage{amssymb}
\usepackage{enumitem}
\usepackage{float}
\usepackage{epsfig}
\usepackage{amsfonts}
\usepackage{latexsym}
\usepackage{amsmath}
\usepackage{mathrsfs}
\usepackage{hyperref}
\usepackage{setspace}
\usepackage{color}
\usepackage{bm}
\usepackage{slashed}
\usepackage{graphicx}  
\usepackage{bm} 
\usepackage{dcolumn}
\usepackage{color}
\usepackage{mathrsfs}
\usepackage{varioref}
\usepackage{hyperref}
\usepackage{dsfont}
\usepackage{caption}
\usepackage{subcaption}
\labelformat{section}{Section #1} 
\labelformat{subsection}{Section #1} 
\labelformat{subsubsection}{Section #1}
\labelformat{subsubsubsection}{Section #1}
\labelformat{equation}{Eq.~(#1)} 
\labelformat{figure}{Fig.~#1} 
\labelformat{subfigure}{Fig.~\thefigure#1} 
\labelformat{table}{Tab.~#1} 
\labelformat{appendix}{Appendix #1}

\title{\bf Emergent $\alpha$-like fermionic vacuum structure and  entanglement in the hyperbolic de Sitter spacetime}
\author{Sourav Bhattacharya\footnote{sbhatta@iitrpr.ac.in}, ~~Shankhadeep Chakrabortty\footnote{s.chakrabortty@iitrpr.ac.in} ~~and ~Shivang Goyal\footnote{2017phz0003@iitrpr.ac.in}\\
{{\small  Department of Physics, Indian Institute of Technology Ropar}}\\ {{\small Rupnagar, Punjab 140 001,
India}}}

\begin{document}
 
\maketitle
\begin{abstract}
\noindent
We report a non-trivial feature of the vacuum structure of  free massive or massless Dirac fields  in the hyperbolic  de Sitter spacetime.  Here we have two causally disconnected regions, say $R$ and $L$ separated by another region, $C$. We are interested in the field theory in $R\cup L$ to understand the long range quantum correlations between $R$ and $L$.  There are local modes of the Dirac field having supports individually either in $R$ or $L$, as well as global modes found via analytically continuing the $R$ modes to $L$ and vice versa. However, we show that unlike the case of a scalar field,   the analytic continuation does not preserve the orthogonality of the resulting global modes. Accordingly, we need to orthonormalise them following the Gram-Schmidt prescription, prior to the field quantisation in order to preserve the canonical anti-commutation relations. We observe that this prescription naturally incorporates   a spacetime independent continuous parameter, $\theta_{\rm RL}$, into the picture. Thus interestingly, we obtain  a naturally emerging one-parameter  family of  $\alpha$-like de Sitter  vacua.   The values of $\theta_{\rm RL}$ yielding the usual thermal spectra of massless created particles are pointed out.
Next, using these vacua, we investigate both entanglement and R\'enyi entropies of either of the regions and demonstrate their dependence on $\theta_{\rm RL}$.  \\
\end{abstract}

\section{Introduction}\label{intro}
The de Sitter  (dS) spacetime is a  maximally symmetric manifold endowed with a positive cosmological constant. It is physically interesting in many ways.  First, owing to the recent observed phase of the accelerated cosmic expansion, there is a strong possibility that our current universe is dominated by a small but positive cosmological constant at large scales. Second, the high degree of homogeneity and isotropy of the current  universe at large scales indicate that our early universe went through an inflationary phase described by a quasi de Sitter spacetime~\cite{Weinberg:2008zzc}.  Being accelerated with expansion and endowed with a cosmological event horizon, the dS offers interesting thermal and other field theoretic properties, we refer our reader to e.g.,~\cite{Gibbons:1977mu, Bhattacharya:2018ltm, Lochan:2018pzs, Higuchi:2018tuk, Solodukhin:2011gn} and also references therein.

It is interesting to investigate the long range quantum correlations between two observers in the dS space, not only causally separated but so by a large distance, say of the order of the superhorizon size. This issue was first addressed in~\cite{Maldacena:2012xp} for a scalar field theory using the coordinatisation of~\cite{Sasaki:1994yt, Bucher:1994gb}, known as the hyperbolic or open chart describing  two casually disconnected expanding regions in dS, as denoted by regions $R$ and $L$ in \ref{fig1}. Since $R$ and $L$ are separated by an entire causally disconnected region $C$, the framework described by \ref{fig1} offers a very natural stage to investigate such long range non-local quantum correlations.  Being motivated by this,  we wish to  investigate in the following the entanglement properties  of the Dirac fermionic vacua in the hyperbolic dS.

Let us first briefly review the case of   a real scalar field~\cite{Maldacena:2012xp}. One first defines orthonormal local basis mode functions  having supports either in $R$ or in $L$, with definite positive or negative frequency behaviour in the asymptotic past. One makes a field expansion using them and  defines the local vacuum as a direct product between the vacua of $R$ and $L$. However, if there is any correlation between these local states, clearly there must exist some mode functions having supports in {\it both } $R$ and $L$. Such {\it global} modes are obtained  by analytically continuing the local modes  from one region into the another along a complex path going through $C$~\cite{Sasaki:1994yt, Bucher:1994gb}.  One then  makes a field quantisation using the global modes and defines a suitable global vacuum. The field quantisations in terms of  the local and global modes give a Bogoliubov relation, yielding in turn an expansion of the global vacuum  in terms of the local states.  It follows that the states belonging to $R$ and $L$ are entangled. The entanglement entropy density is computed using the reduced density operator, found by tracing out the states belonging to either of the regions. Being originated from the long range correlations, the entanglement entropy thus found will not be proportional to any area. This procedure will be more explicit in the due course of the discussion.  

A lot of effort has been given to explore  quantum entanglement in dS so far, in various coordinatisations,  e.g.~\cite{Kanno:2014lma}-\cite{Choudhury:2017qyl}. They not only involve the computations of the entanglement or the R\'enyi entropy (e.g.~\cite{Klebanov:2011uf} and references therein), but also studies of other measures like Bell's inequality, entanglement negativity  and discord, in the Bunch-Davies or more general $\alpha$-vacua (e.g.~\cite{deBoer:2004nd, Collins:2004wj, Feng:2012km} and references therein). We  further refer our reader to e.g.~\cite{Ebadi:2015kqa, Narayan:2015vda, Reynolds:2017lwq, Nguyen:2017ggc, Mahajan:2014daa, Maldacena:2015bha, Boyanovsky:2018soy, Choudhury:2018fpj, Kanno:2018cuk} for discussions on various aspects of quantum correlation in dS, including their possible observational consequences.

Most of the references cited above focus on the scalar field theory and discussions on  other spin fields  seem sparse.  
In~\cite{Kanno:2016qcc}, the entanglement properties of a Dirac fermion in the hyperbolic dS was discussed and certain qualitative differences with a scalar field were pointed out. 
 See~\cite{Boyanovsky:2018soy, Fuentes:2010dt, Kwon:2015gaa, Machado:2018lyn} for interesting aspects of fermionic entanglement in  cosmological spacetimes. See also~\cite{Alsing:2006cj, Montero:2011ai} for discussions on fermionic entanglement in the Rindler spacetime.\\

In this work we wish to point out a further qualitative difference of the fermionic field theory  in the hyperbolic dS from that of the scalar~\cite{Maldacena:2012xp}, which seems to have been missed in the earlier literature, as follows. After reviewing the construction  of the local orthonormal modes and the global ones in \ref{s1}, we show in \ref{s5} that those global modes are {\it not} orthogonal, as opposed to the scalar field theory. It follows from a simple and generic result of the canonical quantum field theory (see~\cite{Wald, Blasone} and references therein), that if one attempts to do field quantisation with such non-orthogonal global modes, the resulting Bogoliubov structure would {\it not  } preserve  the desired anti-commutation relations for the creation and annihilation operators corresponding to the field quantisation of the global modes. Thus we need to orthonormalise those global modes before making any sensible field quantisation. We argue in \ref{s5} that such orthonormalisation is {\it never} unique {\it a priori} in the present scenario, giving rise to a continuous, one parameter family of global vacua. In other words, we wish to demonstrate a natural emergence  of de Sitter $\alpha$-like vacua for Dirac fermions in the hyperbolic de Sitter spacetime. This is the main result of this paper. Using such vacua, we investigate next the fermionic entanglement properties in \ref{BVnEE}.

We shall work with the mostly positive signature of the metric in $(3+1)$-dimensions and will set $c=G=\hbar=\kappa_B=1$ throughout. 

\section{The Dirac modes in hyperbolic dS}\label{s1}
In this section we shall review the basic geometry of the hyperbolic dS, the local Dirac modes and their analytic continuation to form the global modes. The detail of the following can be seen in~\cite{Sasaki:1994yt, Bucher:1994gb,  Kanno:2016qcc, Gromes:1974yu, Camporesi:1995fb}.

\begin{figure}[h!]
	\includegraphics[height=5cm]{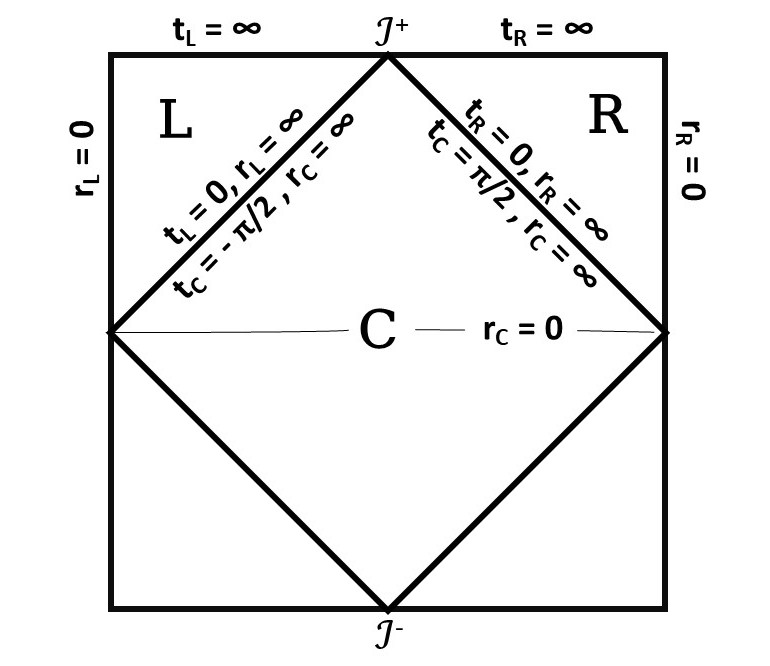}\centering
	\vspace{0.1cm}
	\caption{The Penrose-Carter diagram of the hyperbolic de Sitter spacetime, depicting three causally disconnected regions $R$, $C$ and $L$, e.g.~\cite{Sasaki:1994yt}.  Any $t={\rm const.}$ hypersurface in $R$ or $L$ are $3$-hyperboloids. $R$ and $L$  are superhorizon separated by the intermediate region $C$.  One thus obtains a natural set up to investigate long range quantum correlations between $R$ and $L$. We are only concerned about the expanding $R$ and $L$ regions here.}
	\label{fig1}
\end{figure}

%
The dS  with hyperbolic spatial slicing is obtained by analytic continuation of the Euclidean sphere $S^4$ onto the Lorentzian sector~\cite{Sasaki:1994yt, Bucher:1994gb}. Such analytic continuation gives rise to three distinct and causally disconnected regions of the global dS spacetime, say $R$, $L$ and $C$, as shown in \ref{fig1}. The relationship between the Lorentzian $(t, r)$ and Euclidean coordinates $(\tau, \rho)$ of these three regions is given by, 
\begin{eqnarray}\label{coord}
&&t_R=i\left(\tau-\frac{\pi}{2}\right),   \quad      r_R=i\rho,  \quad  \left(t_R\geq 0, ~~r_R\geq 0\right);\qquad t_L=i\left(-\tau-\frac{\pi}{2}\right),  \quad 
r_L=i\rho,  \quad  \left(t_L\geq 0,~~ r_L\geq 0\right)
\nonumber\\
&&t_C=\tau,  \qquad 
r_C=i\left(\rho-\frac{\pi}{2}\right) \quad  \left(-\frac{\pi}{2}\leq t_C\leq \frac{\pi}{2},~~0\leq r_C< \infty\right). 
\end{eqnarray}
The metrics in the three regions read,
\begin{eqnarray}\label{ee1}
ds^2_{R,L} =H^{-2}\left(
-dt^2_{R,L}+\sinh^2 t_{R,L}(dr_{R,L}^2+\sinh^2 r_{R,L}
d\Omega^2)	\right)
\end{eqnarray}
\begin{eqnarray}\label{ee2}
ds^2_C \ \ \, =H^{-2}\left(
dt^2_C+\cos^2 t_C (-dr_C^2+\cosh^2 r_C d\Omega^2)
\right)
\end{eqnarray}
where $d\Omega^2$ is the metric on ${\cal S}^2$ and $r$ and $t$ are dimensionless in the units of  $H$.

We shall briefly review now the solutions of the free Dirac equation in our  regions of interest $R$ and $L$~\cite{Kanno:2016qcc, Gromes:1974yu, Camporesi:1995fb}. In the region $R$,  the four Dirac modes are given by,
\begin{eqnarray}\label{mode-loc1}
\Psi^{+R}_{(-)}=\left(
\begin{array}{l}
u_{p}(z_R)\\
v_{p}(z_R)
\end{array}\right)
\chi^{(-)}_{pj m}  \qquad  \qquad 
\Psi^{-R}_{(-)}=\left(
\begin{array}{l}
- v_{p}^{\star}(z_R)\\
\ \ u_{p}^{\star}(z_R)
\end{array}\right)
\chi^{(-)}_{pj m} \nonumber\\
\Psi^{-R}_{(+)}=\left(
\begin{array}{l}
\ \ u_{p}^{\star}(z_R)\\
- v_{p}^{\star}(z_R)
\end{array}\right)
\chi^{(+)}_{pj m} \qquad  \qquad 
\Psi^{+R}_{(+)}=
\left(
\begin{array}{l}
v_{p}(z_R)\\
u_{p}(z_R)
\end{array}\right)
\chi^{(+)}_{pj m}. 
\end{eqnarray} 
where $z_R=\cosh t_R$ and a `star' denotes complex conjugation. $\chi^{(\pm)} (r, \theta, \phi)$ are two  orthonormal spatial 2-spinors defined on the 3-hyperboloid, satisfying the eigenvalue equation
\begin{eqnarray}
\widetilde{\slashed\nabla}\chi^{(\pm)}_{p j  m}
=\pm ip\,\chi^{(\pm)}_{p j m}\,,
\label{chi1}
\end{eqnarray}
The `tilde' denotes differentiation over the 3-hyperboloid and since it is an open manifold, the eigenvalue $p$ is continuos and positive.  The temporal parts $u_p$, $v_p$  appearing in \ref{mode-loc1} are given by the hypergeometric functions
\begin{eqnarray}
u_p(z_R)&=&\left(z_R^2-1\right)^{-\frac{3}{4}}\left(\frac{z_R+1}{z_R-1}\right)^{\frac{i p}{2}}
F\left(-\frac{i m}{H},\,\frac{i m}{H},\,\frac{1}{2}-ip,\,\frac{1-z_R}{2}\right)
\label{phi} \nonumber\\
v_p(z_R)&=&-\frac{i m}{H(1-2ip)}\left(z_R^2-1\right)^{-\frac{1}{4}}\left(\frac{z_R+1}{z_R-1}\right)^{\frac{i p}{2}}F\left(1-\frac{i m}{H},\,1+\frac{i m}{H},\,\frac{3}{2}-ip,\,\frac{1-z_R}{2}\right)
\label{varphi}
\end{eqnarray}
The four modes in the region $L$, consistent with \ref{coord} are given by,
\begin{eqnarray}\label{mode-loc2}
&&
\Psi^{+L}_{(-)}(z_L) =
\left(
\begin{array}{r}
v_{p}(z_L)\\
-u_{p}(z_L)
\end{array}\right)
\chi^{(-)}_{pj m}
 \qquad 
 \Psi^{-L}_{(-)}(z_L) =
\left(
\begin{array}{r}
u^\star_{p}(z_L) \\
\ \ v^\star_{p}(z_L) \, \,
\end{array}\right) 
\chi^{(-)}_{pj m} \nonumber \\
&&
\Psi^{-L}_{(+)}(z_L) =
\left(
\begin{array}{r}
\ \  v^\star_{p}(z_L) \, \, \\
u^\star_{p}(z_L) 
\end{array}\right)
\chi^{(+)}_{pj m} \qquad
\Psi^{+L}_{(+)}(z_L) =
\left(
\begin{array}{r}
-u_{p}(z_L)\\
v_{p}(z_L)
\end{array}\right)
\chi^{(+)}_{pj m}
\end{eqnarray}
where as earlier $u_p(z_L)$ and $v_p(z_L)$ are given by \ref{varphi} with $z_L=\cosh t_L$.

It is easy to see that in the asymptotic past, $t_R\to 0$ or $t_L\to 0$, each of the sets \ref{mode-loc1} and \ref{mode-loc2} splits into two positive frequency (representing particles)
and two negative frequency (representing anti-particles) modes. 

The definition of the Dirac inner product between any two modes $\Psi_{(a)}$ and $\Psi_{(b)}$, is given by
\begin{eqnarray}
\left(\Psi_{(a)},\Psi_{(b)}\right):=\int \sinh^3 t\, \sqrt{h}\,  dr d \theta d \phi \,\Psi_{(a)}^{\dagger} \, \Psi_{(b)} 
\label{norm-loc}
\end{eqnarray}
Since the local mode functions have supports only in their respective regions, the sets given in \ref{mode-loc1} and \ref{mode-loc2} are trivially mutually orthogonal.
Also, since the inner product is  independent of time, one can take, in order to simplify the algebra, the Cauchy surface of the integration to be either the $t_R=0$ or the $t_L=0$ hypersurface in the relevant region in \ref{fig1}. Using appropriate simplifications of the hypergeometric functions in this limit~\cite{Copson} and also by using  the  orthonormality of $\chi^{(\pm)}$, it is easy to check that the modes in $R$ or $L$ are orthonormal, i.e.,
\begin{eqnarray} \qquad 
\left(\Psi^{+R}_{(s)}(p,j,m),\,\Psi^{+R}_{(s')}(p',j',m')\right)= \delta(p-p')\, \delta _{j j'}\, \delta_{m m'}\delta_{ss'} 
\quad \left(\Psi^{+R}_{(s)}(p,j,m),\,\Psi^{-R}_{(s')}(p',j',m')\right)= 0~~{\rm etc.} \nonumber\\
\left(\Psi^{+L}_{(s)}(p,j,m),\,\Psi^{+L}_{(s')}(p',j',m')\right)= \delta(p-p')\, \delta _{j j'}\, \delta_{m m'} \delta_{ss'}
\quad \left(\Psi^{+L}_{(s)}(p,j,m),\,\Psi^{-L}_{(s')}(p',j',m')\right)= 0~~{\rm etc.} 
\label{norm-loc'}
\end{eqnarray}
with $s, s' = \pm$ and $(\Psi_{(s)}^{R},\, \Psi_{(s')}^{L})=0$, trivially.

Note also that since we are interested in constructing field theory in $R\cup L$ and the region $C$ is causally disconnected from them, we shall not be concerned about the modes in $C$ in this work.
We further refer our reader to~\cite{Kanno:2016qcc, Yamauchi:2014saa} on the normalisability of the mode functions for Dirac spinors and massive vectors in that region.  \\


In order to understand the quantum correlations of fields located in $R$ and $L$, we also require, apart from the above local modes, the notion of global modes which have support in
$R\cup L$.  First noting that the functions in \ref{varphi} have branch points at $z=\pm 1$ and $\infty$, using  \ref{coord} and some identities of the hypergeometric function~\cite{Copson}, such global modes are achieved by analytically continuing the local modes of $R$ to $L$ or of $L$ to $R$  along a complex path going through $C$~\cite{Kanno:2016qcc}. We review this procedure in \ref{A}. The resulting global mode functions originating from the $R\to L$ analytic continuation are given by,
\begin{eqnarray}\label {globalR}
&&\Psi^{+R}_{(-)}(z_R) =
\left(
\begin{array}{r}
\ \ u_{p}(z_R)   \\
v_{p}(z_R)
\end{array}\right)
\chi^{(-)}_{pj m},\qquad
\Psi^{+R}_{(-)}(z_L) =
\left(
\begin{array}{r}
\lambda_{1}\,v_p(z_L)+\lambda_{2}\,u_{p}^\star(z_L)\\
-\lambda_{1}\,u_p(z_L)+\lambda_{2}\,v_{p}^\star(z_L)
\end{array}\right)
\chi^{(-)}_{pj m}
\nonumber \\
&&\Psi^{+R}_{(+)}(z_R) =
\left(
\begin{array}{r}
\ \ v_{p}(z_R) \, \\
u_{p}(z_R)
\end{array}\right)
\chi^{(+)}_{pj m},\qquad
\Psi^{+R}_{(+)}(z_L) =
\left(
\begin{array}{r}
-\lambda_{1}\,u_p(z_L)+\lambda_{2}\,v_{p}^\star(z_L)\\
\lambda_{1}\,v_p(z_L)+\lambda_{2}\,u_{p}^\star(z_L)
\end{array}\right)
\chi^{(+)}_{pj m}
\nonumber \\
&&\Psi^{-R}_{(-)}(z_R) =
\left(
\begin{array}{r}
-v^\star_{p}(z_R)\\
u^\star_{p}(z_R)
\end{array}\right)
\chi^{(-)}_{pj m},\qquad
\Psi^{-R}_{(-)}(z_L) =
\left(
\begin{array}{r}
 \ \ \lambda_{1} u^\star_p(z_L)-\lambda_{2}^\star v_{p}(z_L) \, \\
\lambda_{1}v^\star_p(z_L)+\lambda_{2}^\star u_{p}(z_L)
\end{array}\right)
\chi^{(-)}_{pj m}
\nonumber \\
&&\Psi^{-R}_{(+)}(z_R) =
\left(
\begin{array}{r}
u^\star_{p}(z_R)\\
- v^\star_{p}(z_R)
\end{array}\right)
\chi^{(+)}_{pj m},\qquad
\Psi^{-R}_{(+)}(z_L) =
\left(
\begin{array}{r}
\ \ \lambda_{1} v^\star_p(z_L)+\lambda_{2}^\star u_{p}(z_L) \, \, \\
\lambda_{1} u^\star_p(z_L)-\lambda_{2}^\star v_{p}(z_L)
\end{array}\right)
\chi^{(+)}_{pj m}
\end{eqnarray}
where
\begin{eqnarray}
\lambda_{1}
=\frac{\sinh\frac{m \pi}{H}}{\cosh\pi p},\qquad
\lambda_{2}=\frac{e^{-\pi p}\,
	\left(\Gamma\left(\frac{1}{2}-ip\right)\right)^2}{\Gamma\left(\frac{1}{2}-ip-\frac{i m}{H}\right)\,
	\Gamma\left(\frac{1}{2}-ip+\frac{i m}{H}\right)}.
\label{AB}
\end{eqnarray}
Likewise we have for the $L\to R$ analytic continuation,
\begin{eqnarray}\label {globalL}
&&
\Psi^{+L}_{(-)}(z_L) =
\left(
\begin{array}{r}
v_{p}(z_L)\\
-u_{p}(z_L)
\end{array}\right)
\chi^{(-)}_{pj m},
\qquad
\Psi^{+L}_{(-)}(z_R) =
\left(
\begin{array}{r}
-\lambda_{1}\,u_p(z_R)+\lambda_{2}\,v_{p}^\star(z_R) \\
-\lambda_{1}\,v_p(z_R)-\lambda_{2}\,u_{p}^\star(z_R)
\end{array}\right)
\chi^{(-)}_{pj m}
\nonumber \\
&&\Psi^{+L}_{(+)}(z_L) =
\left(
\begin{array}{r}
-u_{p}(z_L)\\
v_{p}(z_L)
\end{array}\right)
\chi^{(+)}_{pj m}, 
\qquad
\Psi^{+L}_{(+)}(z_R) =
\left(
\begin{array}{r}
-\lambda_{1}\,v_p(z_R)-\lambda_{2}\,u_{p}^\star(z_R) \\
-\lambda_{1}\,u_p(z_R)+\lambda_{2}\,v_{p}^\star(z_R)
\end{array}\right)
\chi^{(+)}_{pj m}
\nonumber \\
&&\Psi^{-L}_{(-)}(z_L) =
\left(
\begin{array}{r}
u^\star_{p}(z_L) \\
\ \ v^\star_{p}(z_L) \, \,
\end{array}\right)
\chi^{(-)}_{pj m},
\qquad
\Psi^{-L}_{(-)}(z_R) =
\left(
\begin{array}{r}
\lambda_{1}v^\star_p(z_R)+\lambda_{2}^\star u_{p}(z_R) \\
 -\lambda_{1} u^\star_p(z_R)+\lambda_{2}^\star v_{p}(z_R)   \, \,
\end{array}\right)
\chi^{(-)}_{pj m}
\nonumber \\
&&
\Psi^{-L}_{(+)}(z_L) =
\left(
\begin{array}{r}
\ \  v^\star_{p}(z_L) \, \, \\
u^\star_{p}(z_L) 
\end{array}\right)
\chi^{(+)}_{pj m}, \qquad
\Psi^{-L}_{(+)}(z_R) =
\left(
\begin{array}{r}
\ -\lambda_{1} u^\star_p(z_R)+\lambda_{2}^\star v_{p}(z_R)  \ \\
\lambda_{1}v^\star_p(z_R)+\lambda_{2}^\star u_{p}(z_R)
\end{array}\right)
\chi^{(+)}_{pj m}
\end{eqnarray}
Since each of the above mode functions have supports in both the regions, when normalised, they are supposed to be the global versions of the local modes appearing in \ref{mode-loc1}, \ref{mode-loc2}. However, we shall see below that the modes of \ref{globalR} and \ref{globalL} {\it do not} form an orthogonal set under the global inner product, in contrast to the scalar field theory~\cite{Sasaki:1994yt}. As we have discussed in \ref{intro}, we cannot simply treat such non-orthogonal  modes as our global basis modes in the field quantisation~\cite{Wald, Blasone}. Hence we first need to orthonormalise the modes of \ref{globalR}, \ref{globalL}.

\section{Constructing  the global orthonormal modes}\label{s5}

The normalisation integration for the global modes looks formally similar to that of the local ones~\ref{norm-loc}, with the difference that the  integration hypersurface now must exist  in $R\cup L$. Following~\cite{Sasaki:1994yt}, we choose it to be  the $(t_L=0) \cup (t_R=0)$ Cauchy surface  for convenience. Then e.g., for  the pair $\{\Psi^{+L}_{(-)}$, $\Psi^{-R}_{(-)}\}$ we have,
 $$\left( \Psi^{+L}_{(-)},\, \Psi^{-R}_{(-)}\right)_{\rm G} =  \left(\Psi^{+L}_{(-)},\, \Psi^{-R}_{(-)}\right)_{z=z_L=1}+\left(\Psi^{+L}_{(-)},\, \Psi^{-R}_{(-)}\right)_{z=z_R=1}
$$
The suffix `G' stands for global, whereas the inner products on the right hand side are local. We can express  $\Psi^{-R}_{(-)}$ in terms of $z_L$ (for the first term on the right hand side) and  $\Psi^{+L}_{(-)}$ in terms of $z_R$ (for the second term on the right hand side) respectively via \ref{globalR} and \ref{globalL}. Then using \ref{norm-loc'}, we find  (after suppressing the various $\delta$-functions for the sake of brevity)
 $$\left(\Psi^{+L}_{(-)},\, \Psi^{-R}_{(-)}\right)_{\rm G} = -2\lambda_2^{\star}$$
where $\lambda_2$ is given by \ref{AB}.
 It can further be checked in the similar manner  that the eight global fermionic modes of \ref{globalR}, \ref{globalL} can be grouped into four pairs such that the members of any pair are {\it not} orthogonal (although inter-pair orthogonality is satisfied) with respect to the global inner product, given by 
\begin{eqnarray}\label{pair}
\left(\Psi^{+L}_{(-)},\, \Psi^{-R}_{(-)} \right)_{\rm G} =\left(\Psi^{+L}_{(+)},\, \Psi^{-R}_{(+)} \right)_{\rm G}  =-\left(\Psi^{+R}_{(-)}, \, \Psi^{-L}_{(-)} \right)_{\rm G}=-\left(\Psi^{+R}_{(+)}, \,\Psi^{-L}_{(+)} \right)_{\rm G} \,=\, -2\lambda_2^{\star}\, \neq\, 0
\end{eqnarray}
Thus evidently  we cannot use these modes  as our global basis modes, for they would lead to a non-preservation of the canonical anti-commutation relations when used as basis of field expansion, e.g.~\cite{Wald, Blasone}. Hence we need to make suitable linear combinations between  the members of each pair of \ref{pair}, to orthogonalise them via standard Gram-Schmidt procedure. 
  
 Now, we note  from \ref{globalR}, \ref{globalL} that in doing so, we are basically superposing  positive and negative frequency modes. 
Thus in order to accommodate sufficient generality {\it a priori} in our orthogonalisation scheme, we must treat
both the solutions  simultaneously in an {\it equal} footing. 
This can be achieved
by introducing a continuous parametrisation to obtain two orthogonal global modes,
\begin{eqnarray}
\widetilde{\Psi}_{\theta_{\rm RL}} := \Psi^{-R}_{(-)} + \dfrac{2 \lambda_{2}\, \Delta\theta_1}{N_b^2}   \ \Psi^{+L}_{(-)}   \qquad 
\Psi'_{\theta_{\rm RL}}:=  \Psi^{+L}_{(-)} + \dfrac{2 \lambda_{2}^{\star}\, \Delta\theta_2}{N_b^2}   \ \Psi^{-R}_{(-)}
\label{newmodes}
\end{eqnarray}      
where $\Delta\theta_1$ and $\Delta\theta_2$ are one-parameter functions given by
\begin{eqnarray}\label{theta}
\Delta\theta_1 = \dfrac{\cos^2 \theta_{\text{RL}}}{1 - \frac{2 |\lambda_{2}|}{N_b^2} \sin^2\theta_{\text{RL}}} \qquad \qquad \Delta\theta_2 = \dfrac{\sin^2 \theta_{\text{RL}}}{1 + \frac{2 |\lambda_{2}|}{N_b^2} \cos^2\theta_{\text{RL}}} \qquad  \qquad \left(0\leq \theta_{\rm RL} \leq \frac{\pi}{2}\right)
\end{eqnarray}   
and
$$N_b^2 =  \left(1 + \lambda_{1}^2 + |\lambda_{2}|^2\right)$$
 The `angle' $\theta_{\rm RL}$ does not depend upon any spacetime coordinate. It is easy to check that the two mode functions defined in \ref{newmodes} are indeed orthogonal under the global inner product.

In the same spirit, by choosing appropriate linear combinations for the three other pairs in \ref{pair} one can generate orthogonal pairs of global mode functions.  The full set of eight orthonormal global modes is given by, 
\begin{eqnarray}\label{thetamodes}
&&\Psi_{1} = \dfrac{1}{N_1\, N_b} \left( \Psi^{+R}_{(-)} - \dfrac{2 \lambda_{2}  \Delta\theta_1 }{N_b^2}  \, \Psi^{-L}_{(-)} \right)  \qquad 
\Psi_{2} = \dfrac{1}{N_2\,N_b} \left( \Psi^{+R}_{(+)} - \dfrac{2 \lambda_{2}  \Delta\theta_2}{N_b^2} \,   \Psi^{-L}_{(+)} \right) \nonumber\\
&&\Psi_{3} = \dfrac{1}{N_1\,N_b} \left( \Psi^{+L}_{(-)} + \dfrac{2 \lambda_{2}  \Delta\theta_1}{N_b^2} \,    \Psi^{-R}_{(-)} \right) \qquad 
\Psi_{4} = \dfrac{1}{N_2\,N_b} \left( \Psi^{+L}_{(+)} + \dfrac{2 \lambda_{2}   \Delta\theta_2}{N_b^2}   \, \Psi^{-R}_{(+)} \right)  \nonumber\\
&&\Psi_{5} = \dfrac{1}{N_1\,N_b} \left( \Psi^{-R}_{(+)} + \dfrac{2 \lambda_{2}^{\star}  \Delta\theta_1}{N_b^2} \,   \Psi^{+L}_{(+)} \right)  \qquad 
\Psi_{6} = \dfrac{1}{N_2\,N_b} \left( \Psi^{-R}_{(-)} + \dfrac{2 \lambda_{2}^{\star}  \Delta\theta_2}{N_b^2} \,  \Psi^{+L}_{(-)} \right)  \nonumber\\
&&\Psi_{7} = \dfrac{1}{N_1\,N_b} \left( \Psi^{-L}_{(+)} - \dfrac{2 \lambda_{2}^{\star}  \Delta\theta_1 }{N_b^2} \, \Psi^{+R}_{(+)} \right)  \qquad
\Psi_{8} = \dfrac{1}{N_2\,N_b} \left( \Psi^{-L}_{(-)} - \dfrac{2 \lambda_{2}^{\star}  \Delta\theta_2}{N_b^2} \,    \Psi^{+R}_{(-)} \right)  
\end{eqnarray}
where we have written for the normalisations,
\begin{eqnarray}\label{normalisation}
N_1^2 = 1 + \dfrac{4 |\lambda_{2}|^2 \Delta\theta_1^2}{N_b^4}  - \dfrac{8 |\lambda_{2}|^2   \Delta\theta_1}{N_b^4} \qquad
N_2^2 = 1 + \dfrac{4 |\lambda_{2}|^2  \Delta\theta_2^2}{N_b^4}  - \dfrac{8 |\lambda_{2}|^2  \Delta\theta_2}{N_b^4} 
\end{eqnarray}
Using \ref{pair}, the orthonormality of these modes under the global inner product can explicitly be checked at once. To the best of our knowledge, the above issue of orthonormalisation for the fermionic global modes in hyperbolic dS was not addressed in the earlier literature.

Note that the quantisation of the Dirac field with the modes of \ref{thetamodes} will effectively thus give  de Sitter $\alpha$-vacua like structure (see e.g.~\cite{deBoer:2004nd, Collins:2004wj, Feng:2012km, Ashoorioon:2014nta}).  However, we emphasise that unlike the usual cases of  such vacua, introducing the parametrisation $\theta_{\rm RL}$ was  {\it a priori necessary} in our current scenario, in order to maintain sufficient generality in the orthogonalisation procedure.  This is the main result of this work. We shall see later that $\theta_{\rm RL}=0, \pi/2$ values correspond to the usual thermal distribution of created massless particles in $R$ or $L$.

\section{Computation of the entanglement and the R\'enyi entropies}\label{BVnEE}

\subsection{The field quantisation and the Bogoliubov coefficients}\label{bv}
Let us first make a field quantisation in terms of the local modes of \ref{mode-loc1}, \ref{mode-loc2},
\begin{eqnarray}
\label{localdirac}
\Psi=\int dp\sum_{j m s}
\left( c_{(s) p j m}^R\,\Psi^{+R}_{(s) p j m}+d_{(s) p j m}^{R\dag}\,\Psi^{-R}_{(s) p j m}
+c_{(s) p j m}^L\,\Psi^{+L}_{(s) p j m}+d_{(s) p j m}^{L\dag}\,\Psi^{-L}_{(s) p j m} \right),
\end{eqnarray}
where $s =\pm $. The operators are postulated to satisfy the anti-commutation relations
\begin{eqnarray}
\label{antilocal}
&&\left[c_{(s) p j m}^R,~c_{(s') p' j' m'}^{R \dag}\right]_+
=\left[d_{(s) p j m}^R,~d_{(s') p' j' m'}^{R \dag}\right]_+
=\delta\left(p-p'\right)\delta_{ss'}\delta_{jj'}\delta_{mm'} \nonumber\\
&&\left[c_{(s) p j m}^L,~c_{(s') p' j' m'}^{L \dag}\right]_+
=\left[d_{(s) p j m}^L,~d_{(s') p' j' m'}^{L \dag}\right]_+
=\delta\left(p-p'\right)\delta_{ss'}\delta_{jj'}\delta_{mm'}
\end{eqnarray}
    and {\it all} other anti-commutators vanish.  We  define the local vacua $  |0\rangle_R,  |0\rangle_L$, 
\begin{eqnarray}
\label{localvacuua}
c_{(s) p j m}^R  |0\rangle_R = d_{(s) p j m}^R   |0\rangle_R = 0,  \qquad c_{(s) p j m}^L   |0\rangle_L = d_{(s) p j m}^L  |0\rangle_L = 0.
\end{eqnarray}

Likewise, we can also expand the Dirac field in terms of the orthonormal global modes, \ref{thetamodes}.  How do we identify the creation and annihilation operators here? Recalling scalar field's case~\cite{Sasaki:1994yt}, we note from \ref{AB} that in the limit  $p \to \infty$,  both $\lambda_1$ and $\lambda_{2}$ are vanishing, showing  in this limit we do not have any analytically continued modes in \ref{globalR} and \ref{globalL} and  accordingly, \ref{thetamodes} reduces to the local mode functions, having well defined positive or negative energy characteristic in the asymptotic past. Since in that limiting scenario we do not have any trouble with identifying the creation and annihilation operators, we may make the following expansion of the field $\Psi$
in terms of the global modes,
\begin{eqnarray}
\label{globaldirac}
\Psi=\int dp\sum_{j m}
\left( a_{1 p j m}\Psi_{1 p j m} +  a_{ 2 p j m} \Psi_{2 p j m} + a_{ 3 p j m} \Psi_{3 p j m} + a_{ 4 p j m} \Psi_{4 p j m} \right. \nonumber\\ \left. + b_{ 1 p j m}^{\dag} \Psi_{5 p j m}  + b_{ 2 p j m}^{\dag}\Psi_{6 p j m} + b_{ 3 p j m}^{\dag} \Psi_{7 p j m}+ b_{ 4 p j m}^{\dag} \Psi_{8 p j m} \right)
\end{eqnarray}
where we interpret  $a_{1}, \dots, a_4$ and $b_{1}, \dots, b_4$  as the  annihilation operators related to the global modes. The global vacuum is defined as
\begin{eqnarray}
\label{globalvac}
a_{\sigma p j m} |0\rangle = b_{\sigma p j m} |0\rangle =0, \qquad (\sigma = 1,\,2,\,3,\,4)
\end{eqnarray}

We now equate \ref{localdirac} and \ref{globaldirac}, and successively take eight inner products with both sides with respect to the eight global basis modes, \ref
{thetamodes}. While doing so, we need to use \ref{globalR} or \ref{globalL} in order to express the global modes in terms of the local ones.  For the global mode $\Psi_{1 p j m}$ for example, we obtain
\begin{eqnarray}
a_1 &=& \dfrac{1}{N_1 N_b}   \left[ \left(1 -  \alpha^{\star}  \lambda_{2} \right)  c_{(-)}^R  +    \lambda_{1}  c_{(-)}^L  +    \alpha^{\star}  \lambda_{1}  d_{(-)}^{R \dagger}  +   \left( \lambda_{2}^{\star} - \alpha^{\star} \right)  d_{(-)}^{L \dagger}  \right] 
\label{a1}
\end{eqnarray}
where we have written 
$$\alpha = \dfrac{2 \lambda_{2}  \Delta\theta_1}{N_b^2} $$ 
and have suppressed the eigenvalues $p, j, m,$ for the sake of brevity. Similarly, by taking the inner products with the seven other modes $\Psi_2,\,\Psi_3 \dots, \Psi_8$, we obtain
\begin{eqnarray}
a_2 &=& \dfrac{1}{N_2  N_b}   \left[  \left(1 -  \beta^{\star}  \lambda_{2} \right)  c_{(+)}^R  +  \lambda_{1}  c_{(+)}^L  + \beta^{\star}  \lambda_{1}  d_{(+)}^{R \dagger}  +  \left( \lambda_{2}^{\star} - \beta^{\star} \right)  d_{(+)}^{L \dagger}  \right] \nonumber \\
a_3 &=& \dfrac{1}{N_1  N_b}   \left[ - \lambda_{1}  c_{(-)}^R  +  \left(1 -   \alpha^{\star}  \lambda_{2} \right)  c_{(-)}^L  -  \left( \lambda_{2}^{\star} - \alpha^{\star} \right)  d_{(-)}^{R \dagger}  +  \alpha^{\star}  \lambda_{1}   d_{(-)}^{L \dagger}  \right] \nonumber \\
a_4 &=& \dfrac{1}{N_2  N_b}   \left[- \lambda_{1}  c_{(+)}^R  +   \left(1 -  \beta^{\star}  \lambda_{2} \right)  c_{(+)}^L  -  \left( \lambda_{2}^{\star} - \beta^{\star} \right)  d_{(+)}^{R \dagger}  +   \beta^{\star}  \lambda_{1}  d_{(+)}^{L \dagger}  \right] \nonumber \\
b_1^{\dagger} &=& \dfrac{1}{N_1  N_b}   \left[ - \alpha \lambda_{1}  c_{(+)}^R  -  \left(\lambda_{2} - \alpha \right)   c_{(+)}^L  +  \left(1 -  \alpha  \lambda_{2}^{\star} \right)   d_{(+)}^{R \dagger}  +   \lambda_{1}  d_{(+)}^{L \dagger}  \right] \nonumber \\
b_2^{\dagger} &=& \dfrac{1}{N_2  N_b}   \left[  - \beta \lambda_{1}  c_{(-)}^R  -  \left(\lambda_{2} - \beta \right)  c_{(-)}^L  +   \left(1 -  \beta  \lambda_{2}^{\star} \right)   d_{(-)}^{R \dagger}  +   \lambda_{1}  d_{(-)}^{L \dagger}  \right] \nonumber \\
b_3^{\dagger} &=& \dfrac{1}{N_1  N_b}   \left[ \left(\lambda_{2} - \alpha \right)  c_{(+)}^R  -  \alpha \lambda_{1}  c_{(+)}^L  -  \lambda_{1}  d_{(+)}^{R \dagger}  +    \left(1 -  \alpha  \lambda_{2}^{\star} \right)   d_{(+)}^{L \dagger}  \right] \nonumber \\
b_4^{\dagger} &=&  \dfrac{1}{N_2 N_b}   \left[ \left(\lambda_{2} - \beta \right)  c_{(-)}^R  -  \beta \lambda_{1}  c_{(-)}^L  -  \lambda_{1}   d_{(-)}^{R \dagger}  +   \left(1 -  \beta  \lambda_{2}^{\star} \right)   d_{(-)}^{L \dagger}  \right]
\label{ai}
\end{eqnarray}
where we have written
$$ \beta = \dfrac{2 \lambda_{2}   \Delta \theta_2}{N_b^2}$$
Using now \ref{antilocal} it can be checked that the global operators satisfy the desired anti-commutation  relations,
\begin{eqnarray}
\label{antiglobal}
\left[a_{\sigma pj m},~a_{\sigma' p'j' m'}^{\dag}\right]_+
=\left[b_{\sigma p j m},~b_{\sigma' p'j' m'}^{\dag}\right]_+
=\delta\left(p-p'\right)\delta_{\sigma \sigma'}\delta_{jj'}\delta_{mm'},
\end{eqnarray}
where $\sigma= 1,\,2,\,3,\, 4$ with all other anti-commutations vanishing. 

We once again emphasise here that had we not properly orthonormalised our global modes, we would not have retained the above canonical anti-commutation structure, essential to preserve the spin-statistics of the field theory.  

Now, by observing the right hand sides of \ref{a1}, \ref{ai}, it becomes evident that the set of global operators ($a_1$, $a_3$, $b_2$, $b_4$) and ($a_2$, $a_4$, $b_1$, $b_3$) form two disjoint sectors, for their operator contents on the right hand side are different. This implies that the global vacuum defined in \ref{globalvac} can also be split into two subspaces,
\begin{eqnarray}
|0\rangle=|0\rangle^{(1)} \otimes |0\rangle^{(2)}
\end{eqnarray}
where $|0\rangle^1$ is defined by $ a_2 |0\rangle^{(1)}=a_4 |0\rangle^{(1)}=b_1 |0\rangle^{(1)}=b_3|0\rangle^{(1)} = 0$, whereas  $|0\rangle^{(2)}$ is such that $ a_1 |0\rangle^{(2)}=a_3 |0\rangle^{(2)}=b_2 |0\rangle^{(2)}=b_4 |0\rangle^{(2)} = 0$.
For the rest of this paper, we shall work with $|0\rangle^{(1)}$ only. As long as we are only concerned with the computation of the entanglement and R\'enyi entropies, the other subspace,
$|0\rangle^{(2)}$, will produce identical results. 

From \ref{a1}, \ref{ai}, it is clear that  $|0\rangle^{(1)}$ can be constructed as a squeezed state over the local vacua defined in  \ref{localvacuua},
\begin{eqnarray}
|0\rangle^{(1)} = N
\exp\left(\sum_{i,j=R,L} \xi_{ij}\,c^{i\dagger}_{(+)} d^{j\dagger}_{(+)} \right)
 |0\rangle_R \otimes |0\rangle_L,
 \label{vac}
\end{eqnarray}
where $\xi_{ij}$'s are four complex numbers and $N$ is the normalisation.  Also, since the operators $c$'s and $d$'s anti-commute (\ref{antilocal}), we may further decompose the vacua defined in \ref{localvacuua} as,
$$ |0\rangle_R= |0_c\rangle_R \otimes |0_d\rangle_R   \qquad |0\rangle_L=|0_c\rangle_L \otimes |0_d\rangle_L$$   
where $|0_c\rangle_R,~ |0_d\rangle_R,~|0_c\rangle_L,~|0_d\rangle_L$ are annihilated respectively by  $c_{(+)}^R,~ d_{(+)}^R,~c_{(+)}^L,~ d_{(+)}^L$.

We now expand the exponential in \ref{vac}   keeping in mind the various anti-commutations 
and then annihilate $|0\rangle^{(1)}$ by $a_2$, $a_4$, $b_1$, $b_3$, to obtain, 
\begin{eqnarray}
\xi_{RR}=\xi_{LL}= \ -\frac{2 \lambda_{1} \lambda_{2}^\star \left(\lambda_{1}^2+2 \left| \lambda_{2}\right|  \cos 2 \theta_{\text{RL}} +\left| \lambda_{2}\right| ^2+1\right)}{4 \left| \lambda_{2}\right| \left(\lambda_{1}^2+1\right)   \cos 2 \theta_{\text{RL}} +\left| \lambda_{2}\right| ^2 \left(2 \lambda_{1}^2+\cos4 \theta_{\text{RL}}+1\right)+2 \left(\lambda_{1}^2+1\right)^2} =\xi_1 ~~({\rm say})\nonumber\\
%
\xi_{RL}=-\xi_{LR}=  -\frac{\lambda_{2}^\star \left(2 \left(\lambda_{1}^2+1\right) \cos 2 \theta_{\text{RL}} +2 \left| \lambda_{2}\right| ^2 \cos 2 \theta_{\text{RL}} +\left| \lambda_{2}\right|  (\cos 4 \theta_{\text{RL}} +3)\right)}{4 | \lambda_{2}|  \left(\lambda_{1}^2+1\right)  \cos 2 \theta_{\text{RL}} +\left| \lambda_{2}\right| ^2 \left(2 \lambda_{1}^2+\cos 4 \theta_{\text{RL}} +1\right)+2 \left(\lambda_{1}^2+1\right)^2}=\xi_2 ~~({\rm say})
\label{xi12}
\end{eqnarray}
Also, the normalisation in \ref{vac} reads,
\begin{eqnarray}
N=  \left( 1 + 2  \left| \xi_{1} \right| ^2 + 2  \left| \xi_{2}\right|^2 +  \left|  \xi_{1}^2 + \xi_{2}^2 \right| \right)^{-1/2}
\label{norm}
\end{eqnarray}
Putting these all in together, we may now explicitly write down the global vacuum in terms of the local states,
\begin{eqnarray}
|0\rangle^{(1)} = N \, \left[ |0_c\rangle_R \otimes |0_d\rangle_R \otimes |0_c\rangle_L  \otimes
 |0_d\rangle_L \,+\, \xi_{1}\, \left\{|1_c \rangle_R \otimes |1_d\rangle_R \otimes |0_c \rangle_L  \otimes 
 |0_d \rangle_L\, +\, |0_c\rangle_R \otimes |0_d\rangle_R \otimes |1_c\rangle_L \otimes |1_d\rangle_L \right\} \right. \nonumber\\ \left.
+\xi_{2}\, \left\{ |1_c \rangle_R \otimes | 0_d\rangle_R  \otimes |0_c\rangle_L \otimes |1_d \rangle_L \,+\, |0_c\rangle_R \otimes |1_d\rangle_R \otimes |1_c\rangle_L \otimes | 0_d\rangle_L \right\}\, +\, (\xi_{1}^2 +  \xi_{2}^2)\,\, |1_c\rangle_R\otimes |1_d\rangle_R \otimes  |1_c\rangle_L \otimes |1_d\rangle_L  \right] \nonumber\\
\label{vac2}
\end{eqnarray}
The above expression shows that there will be pair creation in $R$ and $L$  with respect to the global vacuum. Since the states belonging to $R$ and $L$ cannot be factored out, those pairs will be entangled. Also, since $|0\rangle^{(1)} $ depends upon $\theta_{\rm RL}$ through $\xi_1$ and $\xi_2$, the particle creation and the entanglement will also depend upon it.

 Let us then first compute, as a check of consistency of the entire framework, the expectation value of the local number operator with respect to the global vacuum, which will give us the number of created particles at a given mode.   We shall do it in the massless limit only, for which from \ref{xi12} and \ref{vac2} we have,
\begin{eqnarray}
{}^{(1)}\langle 0 | c_{(+)}^{R \dag} c_{(+)}^{R } | 0 \rangle^{(1)} = N^2 \left|  \xi_{2}\right| ^2 \left(  1+  \left|\xi_{2}\right|^2 \right) = \dfrac{1}{1 + e^{2 p \pi}\left(\frac{e^{p \pi} + \cos 2\theta_{\text{RL}} } {1 + e^{p \pi} \cos 2 \theta_{\text{RL}}}\right)^2 }
\end{eqnarray}
For $\theta_{\text{RL}} \to 0, \pi/2$, we reproduce the familiar  Fermi-Dirac distribution,
$$\frac{1}{e^{2 \pi p } + 1}$$
Away from these values of $\theta_{\rm RL}$, the spectrum is not thermal.  
 For $\alpha$-vacua in the dS spacetime with flat spatial slicing, such non-thermality was also noted earlier in~\cite{Feng:2012km}.

\subsection{The entanglement entropy }\label{ee}
We start with the full density operator $\rho_{\rm global} = |0\rangle^{(1)} \, {}^{(1)}\langle 0|$ and trace over the states of, say  $L$ region, inaccessible to an observer in $R$.  We thus obtain the reduced density operator in $R$,
\begin{eqnarray}
\rho_R = \operatorname{Tr}_L \rho_{\rm global}.
\end{eqnarray}
Tracing over the $R$ states will  give identical results. We find a matrix representation of $\rho_R$ in terms of the $R$-state vectors,
\begin{eqnarray}
\rho_R \equiv |N|^2 \left(
\begin{array}{cccc}
1 + \left| \xi_{1}\right| ^2 & 0 & 0 & \xi_{1}^{\star} + \xi_{1} \left(\xi_{1}^{\star2} +  \xi_{2}^{\star 2} \right) \vspace{3mm}\\
0 & \left|  \xi_{2}\right| ^2 & 0 & 0 \vspace{3mm}\\
0 & 0 & \left|  \xi_{2}\right| ^2 & 0 \vspace{3mm}\\
\xi_{1} + \xi_{1}^{\star} \left(\xi_{1}^2 +  \xi_{2}^2\right) & 0 & 0 & \left| \xi_{1} \right| ^2 + \left|  \xi_{1}^2 +  \xi_{2}^2  \right| ^2  \\
\end{array}\right)
\label{matrix}
\end{eqnarray}
The entanglement entropy for a single mode (with a given $p$ and $\theta_{\rm RL}$ value) is given by
\begin{eqnarray}
S(p, m; \theta_{\rm RL}) = -{\rm Tr}\left(\rho_R\ln\rho_R\right) = -\sum_{i=1}^4 \lambda_R^{(i)} \ln \lambda_R^{(i)}
\end{eqnarray}
where $\lambda_R^{(i)}$'s are the eigenvalues of \ref{matrix}. 
%
%
%
%
%
%
The entanglement entropy per unit volume between $R$ and $L$ is obtained by integrating over all $p$  modes. The final expression of the entanglement entropy is obtained by further integrating the result over the purely spatial section of \ref{ee1}. Since 
 $S(p, m; \theta_{\rm RL})$ has no spatial dependence, the integration, being done over a non-compact space, would diverge. Accordingly, one needs to put a cut off at some `large' radial distance. The resultant regularised volume integral equals  $2\pi$, see~\cite{Maldacena:2012xp} for details. The regularised entanglement entropy then equals  
\begin{eqnarray}
S(m, \theta_{\text{RL}})=2\pi\int_0^\infty dp\,{\cal D}(p) S(p,m, \theta_{\text{RL}})
\label{integrate'}
\end{eqnarray}
where ${\cal D}(p)=(1/4+p^2)/(2\pi^2)$ is the apprpriate measure in the momentum space corresponding to the spatial section of \ref{ee1} for the spin-1/2 field~\cite{Camporesi:1995fb, Bytsenko:1994bc}. The integral in \ref{integrate'} is convergent and can be calculated numerically. 

\begin{figure}[H]
	\includegraphics[height=5.0cm]{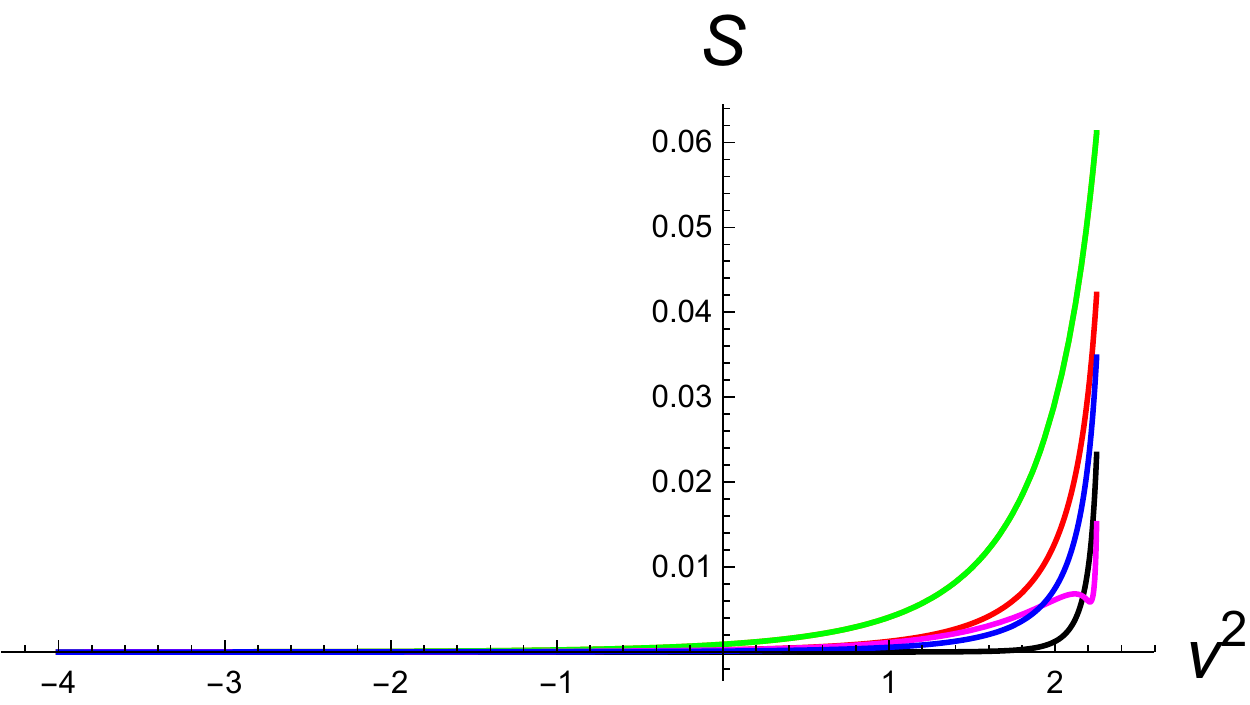}\centering
	\vspace{0.01cm}
	\caption{Entanglement entropy versus $\nu^2=9/4-m^2/H^2$ plots for various values of the parametrisation $\theta_{\text{RL}}$. The green curve corresponds to $\theta_{\text{RL}}  = 0, \,\pi/2$, the red to $\theta_{\text{RL}} = \pi/6$, the blue  to $\theta_{\text{RL}} = \pi/5$, the black to $\theta_{\text{RL}} = \pi/4$ and the pink to $\theta_{\text{RL}} = \pi/3$. We may chiefly note that a) the  $\theta_{\text{RL}}  = 0, \,\pi/2$ plots are exactly coincident and they give maximum entanglement entropy for all values of $\nu^2$  b) the pink curve ($\theta_{\text{RL}} = \pi/3$) has extrema, whereas the other plots are monotonic.  c) for any given value of  $\theta_{\rm RL}$, the entanglement entropy has its maximum value in the massless limit, $\nu^2=9/4$.   }
	\label{plot1}
\end{figure}

\begin{figure}[H]
	\includegraphics[height=5cm]{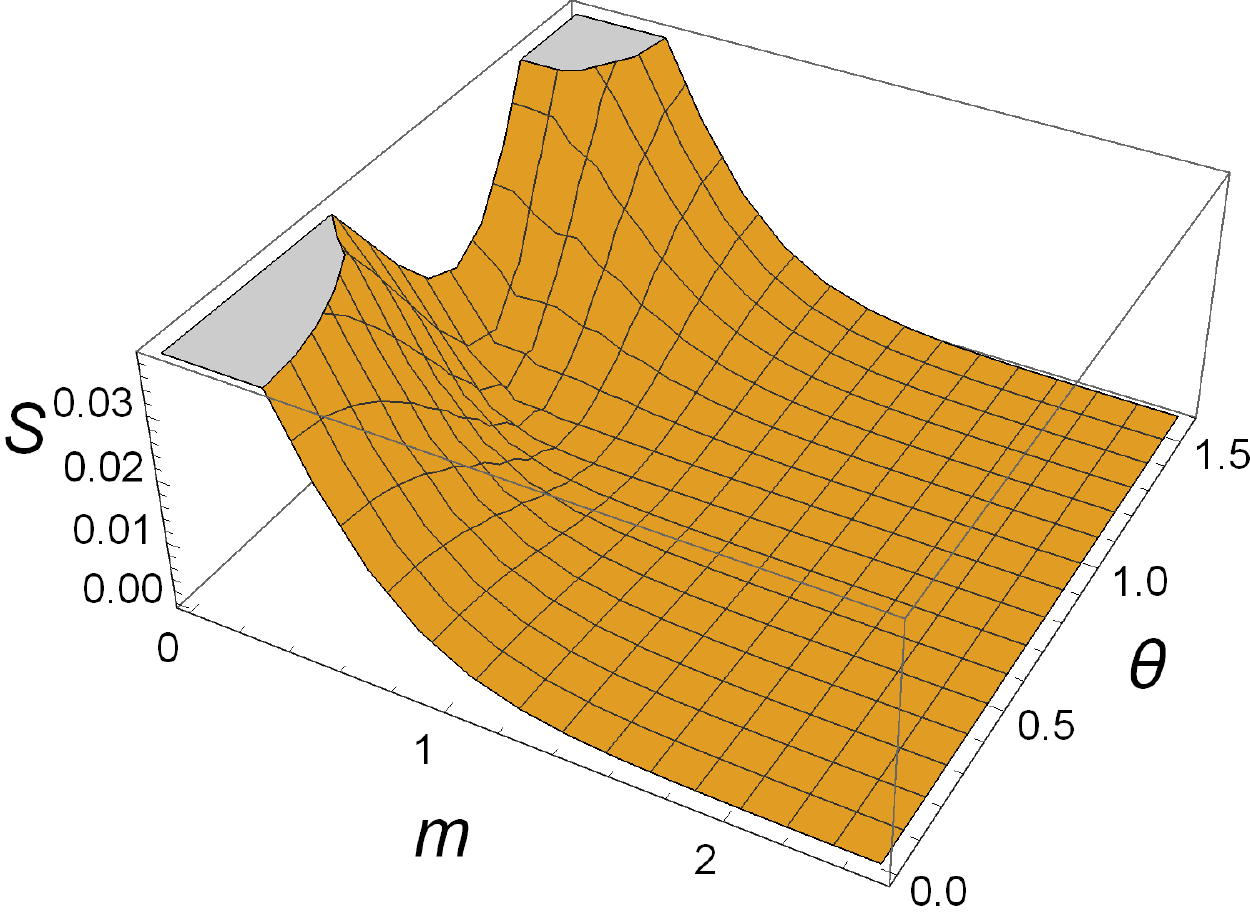}\centering
	\vspace{0.1cm}
	\caption{A three dimensional plot depicting the variation of the entanglement entropy with $\theta_{\text{RL}}$ and the mass $m$ in an equal footing. }
	\label{plot3}
\end{figure}

We have plotted various characteristics of the entanglement entropy in \ref{plot1} and \ref{plot3} as a function of the parameter $\nu^2=9/4-m^2/H^2$ subject to different values of $\theta_{\rm RL}$. We may chiefly note the following features.

a) In \ref{plot1}, the curves corresponding to  $\theta_{\rm RL}=0,\, \pi/2$, are exactly coincident and they show maximal $R-L$ entanglement for all values of $\nu^2$. The coincidence corresponds to the fact  that the coefficients $\xi_{1,2}(\theta_{\rm RL}=0)=\pm \xi_{1,2}(\theta_{\rm RL}=\pi/2)$, in \ref{xi12}. b) While most of the curves in \ref{plot1} are monotonic, the curve corresponding to $\theta_{\rm RL}=\pi/3$ shows extrema. c) For any given value of  $\theta_{\rm RL}$, the entanglement entropy has its maximum value in the massless limit, $\nu^2=9/4$. This might correspond to the fact that in this limit the creation of particle-antiparticle pairs is energetically most favourable. Since such pairs are entangled, \ref{vac2}, it is perhaps reasonable to   expect that the entanglement entropy also gets its maximum value in the massless limit. 

%

 \subsection{The R\'enyi entropy}\label{re}
 Before we end, we wish to briefly discuss  the R\'enyi entropy,  a one-parameter generalisation of entanglement entropy, e.g.~\cite{Klebanov:2011uf}
\begin{eqnarray}
    S_q = \dfrac {\ln {\rm Tr} \rho^q}{1 - q}, \qquad   q>0
\label{Reni}
\end{eqnarray}
For $q\to 1$, \ref{Reni} reduces to the entanglement entropy, as can be easily seen by using the L'Hopital's rule to the above equation.

\begin{figure}[H]
\centering
\begin{subfigure}{.5\textwidth}
	\centering
	\includegraphics[width=0.95\linewidth]{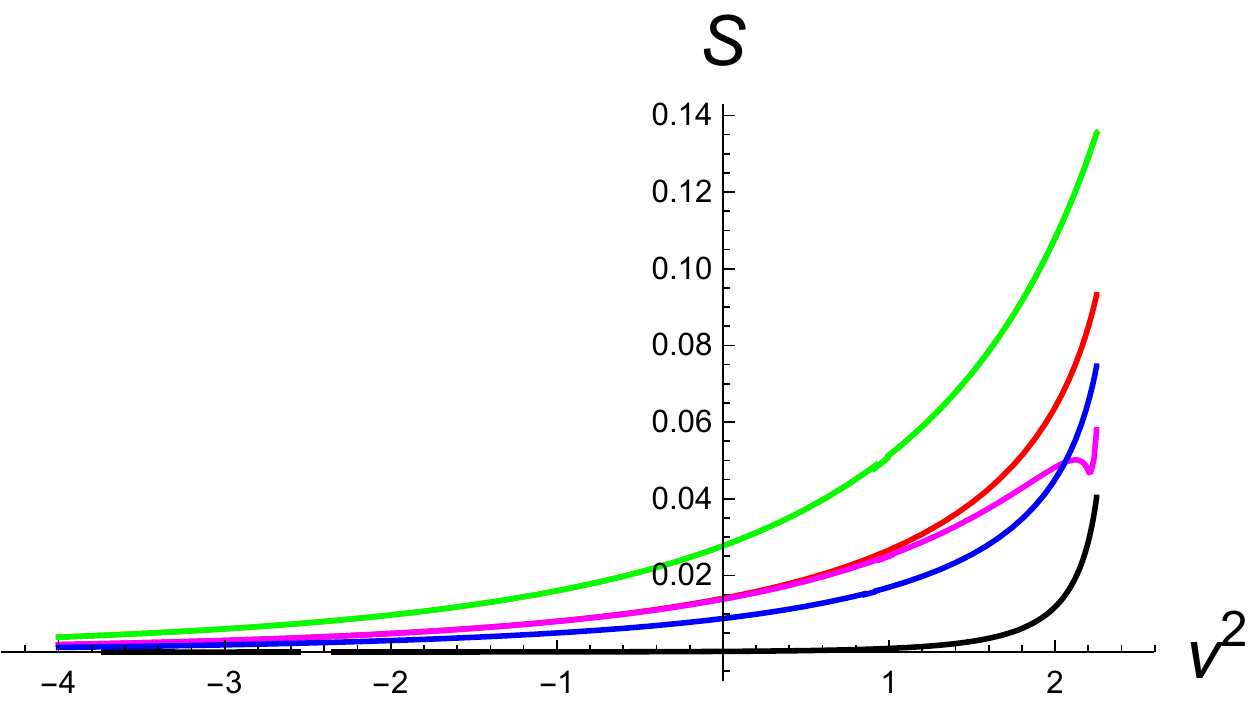}
	\caption{ }
	\label{plot5}
\end{subfigure}%
\begin{subfigure}{.5\textwidth}
	\centering
	\includegraphics[width=0.95\linewidth]{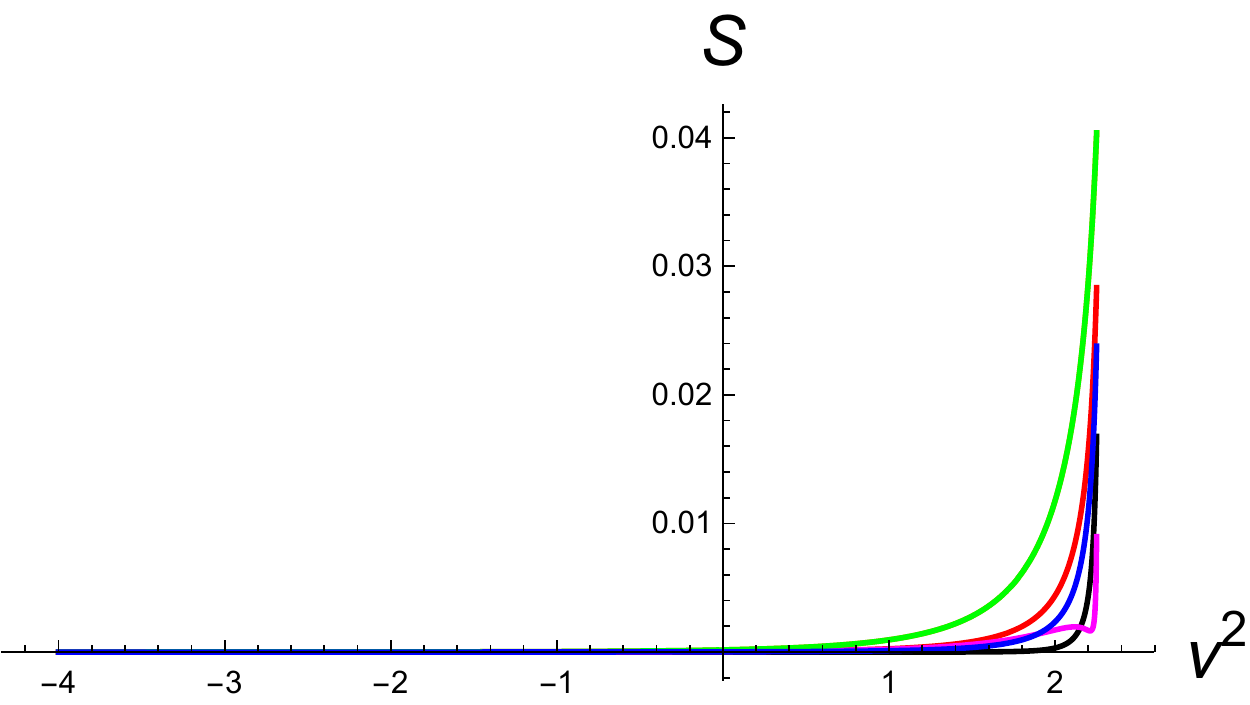}
	\caption{ }
	\label{plot4}
\end{subfigure}
\caption{The R\'enyi entropy for (a) $q = 0.5$ and (b) $q = 2$, for various values of $\theta_{\text{RL}}$. The green curve corresponds to $\theta_{\text{RL}}  = 0, \,\pi/2$, the red to $\theta_{\text{RL}} = \pi/6$, the blue  to $\theta_{\text{RL}} = \pi/5$, the black to $\theta_{\text{RL}} = \pi/4$ and the pink to $\theta_{\text{RL}} = \pi/3$.}
\label{plot4'}
\end{figure}

The final R\'enyi entropy, akin to the expression for the final regularised entanglement entropy, is given by,
\begin{eqnarray}
S_q(m, \theta_{\text{RL}})=2\pi\int_0^\infty dp\,{\cal D}(p) \ S_q(p,m, \theta_{\text{RL}})
\label{integrate}
\end{eqnarray}
We have plotted $S_q(m, \theta_{\text{RL}})$ in \ref{plot5}, \ref{plot4}  as earlier with respect to the parameter $\nu^2=9/4-m^2/H^2$, with different values of $\theta_{\rm RL}$. We note chiefly that as a whole, the qualitative nature of the R\'enyi entropy for different $q$ values  remains the same as that of the entanglement entropy. In particular, a) the values $\theta_{\rm RL}=0, \pi/2$ gives maximum R\'enyi entropy for all values of $\nu^2$ and b) the extrema for $\theta=\pi/3$ is still present.
%

\section{Discussions}\label{disc}
In this paper  we have investigated 
the entanglement and the R\'enyi entropies between the $R$ and $L$ states of the Dirac field in the hyperbolic dS spacetime, \ref{ee1}, \ref{ee2}, as a measure of the long range non-local quantum correlations between these two regions.

The chief results of these paper could be summarised as follows. First, the {\it natural emergence} of the continuous, one parameter family of global modes (cf.~\ref{s5}) and vacua, \ref{vac2}. 
Such vacua have structures similar to the  de Sitter $\alpha$-vacua, though they have originated here from the mere necessity of an {\it a priori} general orthonormalisation scheme for the global modes. Such orthonormalisation of the modes is necessary to preserve the canonical anti-commutation structure of the field theory~\cite{Wald, Blasone}. Second, we have seen in \ref{ee}, \ref{re} that a) $\theta_{\rm RL}=0, \pi/2$ reproduces the thermal spectrum for the created massless particles. The dependence of the entanglement and the R\'enyi entropies on the parametrisation $\theta_{\rm RL}$ was depicted in \ref{plot1}, \ref{plot3}, \ref{plot5}, \ref{plot4}.

We recall that instead of taking $\alpha$ as a usual independent parameter, one can also take it to be momentum dependent, see~\cite{deBoer:2004nd}  for a discussion on the scalar field theory. Note that even though we have not taken any momentum dependence in $\theta_{\rm RL}$ here, the coefficients of the linear combinations in the global modes,~\ref{thetamodes}, are indeed momentum dependent. This effectively makes our construction qualitatively similar to that of~\cite{deBoer:2004nd}. Nevertheless, it will be interesting  on top of this to further allow explicit momentum dependence in $\theta_{\rm RL}$.  In this case we need to make a suitable ansatz for it, such that
the mode by mode normalisability of the states is achieved and also the various momentum integrals we encounter converge. 

It seems interesting to investigate the effects of $\theta_{\rm RL}$
into the other measures of quantum correlations e.g., the entanglement negativity, the violation of Bell's inequality and the quantum discord etc, in order to quantify it further. We hope to address these issues in our future work.

\section*{Acknowledgements}
We would like to acknowledge R.~Basu, S.~Chakraborty, R.~Gupta, A.~Lahiri, S.~Mukherjee, B.~Sathiapalan and I.~S.~Tyagi for fruitful discussions. SB's research is partially supported by the ISIRD grant 9-289/2017/IITRPR/704. SC is partially supported by ISIRD grant 9-252/2016/IITRPR/708. 

\bigskip
\appendix
\labelformat{section}{Appendix #1} 
\section{The analytic continuation and the global modes}\label{A}
Here we review the construction of the global modes found via the analytic continuation of the local mode functions of \ref{mode-loc1}, \ref{mode-loc2},~\cite{Kanno:2016qcc}. 
Since the hypergeometric function has branch points at $1$ and $\infty$, the functions $u_p(z_R)$ and $v_p(z_R)$ in \ref{varphi} have branch points at $\pm 1$ and $\infty$. We join $\pm 1$ with a cut on the real line and further join either of them to $\infty$ in any possible way we wish, ensuring however, that we do not hit the cut to $\infty$ during the  process described  below. The spatial functions $\chi^{(\pm)}$ however, do not undergo any formal changes under this procedure. \\

We divide the real $z$ line into three regions : a) $z \geq 1$ ($z_R$ or the $R$ region) b) $-1\leq z\leq1$ (the $C$ region) and c) $z<-1$. Note that both $z_R$ and $z_L$ appearing in the mode functions are greater than unity. Thus when we analytically continue an $R$ mode to the $L$ region via $C$, we eventually reach at negative $z$ values. Hence while continuing analytically, we shall take $z_R=-z_L$ with $z_L>1$. \\

We now extend $u_p(z_R)$ from
$R$ to $C$ by taking
$z_R-1=e^{-\pi i}(1-z_R)$ and change the variable by $z_R=-z_L$, as mentioned above. 
Since  $z_L>1$, we perform $1-z_L=e^{\pi i}(z_L-1)$, to obtain the continued form of $u_p(z_R)$,
\begin{eqnarray}
\left(e^{- \pi i}\right)^{-\frac{3}{4} - \frac{i p}{2}}  \left(1+z_L\right)^{-\frac{3}{4} - \frac{i p}{2}}   \left(1-z_L\right)^{-\frac{3}{4} + \frac{i p}{2}}
F\left(-\frac{i m}{H}\,,\,\frac{i m}{H}\,,\,\frac{1}{2}-ip\,,\,\frac{1+z_L}{2}\right)
\label{ac1}
\end{eqnarray}
In order to cast the above expression into the form of our initial mode functions, we recall the  identity, e.g.~\cite{Copson}
\begin{eqnarray}
F\left(\alpha,\beta,\gamma;z\right)&=&\frac{\Gamma(\gamma)\Gamma(\alpha+\beta-\gamma)}{\Gamma(\alpha)\Gamma(\beta)}\left(1-z\right)^{\gamma-\alpha-\beta}F\left(\gamma-\alpha,\gamma-\beta,\gamma-\alpha-\beta+1;1-z\right)\nonumber\\
&&+\frac{\Gamma(\gamma)\Gamma(\gamma-\alpha-\beta)}{\Gamma(\gamma-\alpha)\Gamma(\gamma-\beta)}
F\left(\alpha,\beta,\alpha+\beta-\gamma+1;1-z\right),
\end{eqnarray} 
Substituting this into \ref{ac1} with $z\equiv (1+z_L)/2$ and also using 
$$|\Gamma(1+ix)|^2 = \frac{\pi x}{\sinh\pi x}$$
we finally obtain
\begin{eqnarray}
u_p(z_R) = \lambda_{1}\,v_p(z_L)+\lambda_{2}\,u_{p}^\star(z_L)
\label{ac2}
\end{eqnarray}
where $\lambda_1$, $\lambda_2$ are given by \ref{AB}.

Similarly we find,
\begin{eqnarray}
v_p(z_R) = -\lambda_{1}\,u_p(z_L)+\lambda_{2}\,v_{p}^\star(z_L)
\end{eqnarray}
The opposite procedure, i.e. the analytic continuation $L\to R$ yields  formally exactly similar results.

\bigskip


\begin{thebibliography}{99} 

\bibitem{Weinberg:2008zzc} 
  S.~Weinberg,
  {\it Cosmology},
  Oxford, UK: Oxford Univ. Pr. (2008)
  
 
  
  
    
   \bibitem{Gibbons:1977mu} 
     G.~W.~Gibbons and S.~W.~Hawking,
  {\it Cosmological Event Horizons, Thermodynamics, and Particle Creation},
  Phys.\ Rev.\ D {\bf 15}, 2738 (1977)
  
   \bibitem{Bhattacharya:2018ltm} 
  S.~Bhattacharya,
  {\it Particle creation by de Sitter black holes revisited}, to appear in Phys.~Rev.~D,
  arXiv:1810.13260 [gr-qc]
  

  \bibitem{Lochan:2018pzs} 
  K.~Lochan, K.~Rajeev, A.~Vikram and T.~Padmanabhan,
  {\it Quantum correlators in Friedmann spacetimes: The omnipresent de Sitter spacetime and the invariant vacuum noise},
  Phys.\ Rev.\ D {\bf 98}, no. 10, 105015 (2018)
  [arXiv:1805.08800 [gr-qc]]
  
  \bibitem{Higuchi:2018tuk} 
  A.~Higuchi and K.~Yamamoto,
{\it Vacuum state in de Sitter spacetime with static charts},
  Phys.\ Rev.\ D {\bf 98}, no. 6, 065014 (2018)
  [arXiv:1808.02147 [gr-qc]]


\bibitem{Solodukhin:2011gn} 
  S.~N.~Solodukhin,
 {\it Entanglement entropy of black holes},
  Living Rev.\ Rel.\  {\bf 14}, 8 (2011)
  [arXiv:1104.3712 [hep-th]]
  
      
   \bibitem{Maldacena:2012xp} 
   J.~Maldacena and G.~L.~Pimentel,
  {\it Entanglement entropy in de Sitter space},
  JHEP {\bf 1302}, 038 (2013)
  [arXiv:1210.7244 [hep-th]]
  
   \bibitem{Sasaki:1994yt} 
  M.~Sasaki, T.~Tanaka and K.~Yamamoto,
  {\it Euclidean vacuum mode functions for a scalar field on open de Sitter space},
  Phys.\ Rev.\ D {\bf 51}, 2979 (1995)
  [gr-qc/9412025]



\bibitem{Bucher:1994gb} 
  M.~Bucher, A.~S.~Goldhaber and N.~Turok,
  {\it An open universe from inflation},
  Phys.\ Rev.\ D {\bf 52}, 3314 (1995)
  [hep-ph/9411206]
  

  
  \bibitem{Kanno:2014lma} 
  S.~Kanno, J.~Murugan, J.~P.~Shock and J.~Soda,
 {\it Entanglement entropy of $\alpha$-vacua in de Sitter space},
  JHEP {\bf 1407}, 072 (2014)
  [arXiv:1404.6815 [hep-th]]
  
  \bibitem{Iizuka:2014rua} 
N.~Iizuka, T.~Noumi and N.~Ogawa,
{\it Entanglement entropy of de Sitter space $\alpha$-vacua},
Nucl.\ Phys.\ B {\bf 910}, 23 (2016)
[arXiv:1404.7487 [hep-th]]

  
  \bibitem{Kanno:2014ifa} 
  S.~Kanno,
 {\it Impact of quantum entanglement on spectrum of cosmological fluctuations},
  JCAP {\bf 1407}, 029 (2014)
[arXiv:1405.7793 [hep-th]]

\bibitem{Kanno:2014bma}  
  S.~Kanno, J.~P.~Shock and J.~Soda,
{\it Entanglement negativity in the multiverse},
  JCAP {\bf 1503}, 015 (2015)
  [arXiv:1412.2838 [hep-th]]
  
  
  \bibitem{Kanno:2015lja} 
  S.~Kanno,
  {\it Cosmological implications of quantum entanglement in the multiverse},
  Phys.\ Lett.\ B {\bf 751}, 316 (2015)
  [arXiv:1506.07808 [hep-th]]
  
  
  \bibitem{Kanno:2015ewa} 
  S.~Kanno,
 {\it A note on initial state entanglement in inflationary cosmology},
  EPL {\bf 111}, no. 6, 60007 (2015)
  [arXiv:1507.04877 [hep-th]]
  
  \bibitem{Choudhury:2016cso} 
  S.~Choudhury, S.~Panda and R.~Singh,
  {\it Bell violation in the Sky},
  Eur.\ Phys.\ J.\ C {\bf 77}, no. 2, 60 (2017)
  [arXiv:1607.00237 [hep-th]].

\bibitem{Choudhury:2016pfr} 
  S.~Choudhury, S.~Panda and R.~Singh,
 {\it Bell violation in primordial cosmology},
  Universe {\bf 3}, no. 1, 13 (2017)
  [arXiv:1612.09445 [hep-th]].


  \bibitem{Kanno:2016gas} 
  S.~Kanno, J.~P.~Shock and J.~Soda,
  {\it Quantum discord in de Sitter space},
  Phys.\ Rev.\ D {\bf 94}, no. 12, 125014 (2016)
  [arXiv:1608.02853 [hep-th]]

\bibitem{Vancea:2016tkt} 
I.~V.~Vancea,
{\it Entanglement Entropy in the $\sigma$-Model with the de Sitter Target Space},
Nucl.\ Phys.\ B {\bf 924}, 453 (2017)
[arXiv:1609.02223 [hep-th]]




\bibitem{Soda:2017yzu} 
  J.~Soda, S.~Kanno and J.~P.~Shock,
  {\it Quantum Correlations in de Sitter Space},
  Universe {\bf 3}, no. 1, 2 (2017)

\bibitem{Kanno:2017wpw} 
  S.~Kanno,
 {\it Quantum Entanglement in the Multiverse},
  Universe {\bf 3}, no. 2, 28 (2017)

\bibitem{Kanno:2017dci} 
  S.~Kanno and J.~Soda,
{\it Infinite violation of Bell inequalities in inflation},
  Phys.\ Rev.\ D {\bf 96}, no. 8, 083501 (2017)
  [arXiv:1705.06199 [hep-th]]
  
  

\bibitem{Holland:2017cza} 
  J.~Holland, S.~Kanno and I.~Zavala,
{\it Anisotropic Inflation with Derivative Couplings},
  Phys.\ Rev.\ D {\bf 97}, no. 10, 103534 (2018)
  [arXiv:1711.07450 [hep-th]]
  
  \bibitem{Albrecht:2018prr} 
  A.~Albrecht, S.~Kanno and M.~Sasaki,
{\it Quantum entanglement in de Sitter space with a wall, and the decoherence of bubble universes},
  Phys.\ Rev.\ D {\bf 97}, no. 8, 083520 (2018)
  [arXiv:1802.08794 [hep-th]]
  
  
  
\bibitem{Choudhury:2018rjl} 
S.~Choudhury, A.~Mukherjee, P.~Chauhan and S.~Bhattacherjee,
{\it Quantum Out-of-Equilibrium Cosmology},
arXiv:1809.02732 [hep-th]

\bibitem{Feng:2018ebt} 
J.~Feng, X.~Huang, Y.~Z.~Zhang and H.~Fan,
{\it Bell inequalities violation within non-Bunch-Davies states},
Phys.\ Lett.\ B {\bf 786}, 403 (2018)
[arXiv:1806.08923 [hep-th]]



\bibitem{Choudhury:2017bou} 
S.~Choudhury and S.~Panda,
{\it Entangled de Sitter from stringy axionic Bell pair I: an analysis using Bunch-Davies vacuum},
Eur.\ Phys.\ J.\ C {\bf 78}, no. 1, 52 (2018)
[arXiv:1708.02265 [hep-th]]



\bibitem{Choudhury:2017qyl} 
S.~Choudhury and S.~Panda,
{\it Quantum entanglement in de Sitter space from Stringy Axion: An analysis using $\alpha$ vacua},
arXiv:1712.08299 [hep-th]

\bibitem{Klebanov:2011uf} 
  I.~R.~Klebanov, S.~S.~Pufu, S.~Sachdev and B.~R.~Safdi,
  {\it Renyi Entropies for Free Field Theories},
  JHEP {\bf 1204}, 074 (2012)
  [arXiv:1111.6290 [hep-th]]
  
  \bibitem{deBoer:2004nd} 
  J.~de Boer, V.~Jejjala and D.~Minic,
 {\it Alpha-states in de Sitter space},
  Phys.\ Rev.\ D {\bf 71}, 044013 (2005)
  [hep-th/0406217]
  
\bibitem{Collins:2004wj} 
  H.~Collins,
{\it Fermionic alpha-vacua},
  Phys.\ Rev.\ D {\bf 71}, 024002 (2005)
  [hep-th/0410229]
  
  
  
  \bibitem{Feng:2012km} 
  J.~Feng, Y.~Z.~Zhang, M.~D.~Gould, H.~Fan, C.~Y.~Sun and W.~L.~Yang,
  {\it Probing Planckian physics in de Sitter space with quantum correlations},
  Annals Phys.\  {\bf 351}, 872 (2014)
  [arXiv:1211.3002 [quant-ph]]
  

\bibitem{Ebadi:2015kqa} 
Z.~Ebadi and B.~Mirza,
{\it Entanglement generation due to the background electric field and curvature of space-time},
Int.\ J.\ Mod.\ Phys.\ A {\bf 30}, no. 07, 1550031 (2015)

\bibitem{Narayan:2015vda} 
K.~Narayan,
{\it Extremal surfaces in de Sitter spacetime},
Phys.\ Rev.\ D {\bf 91}, no. 12, 126011 (2015)
[arXiv:1501.03019 [hep-th]]

\bibitem{Reynolds:2017lwq} 
  A.~Reynolds and S.~F.~Ross,
  {\it Complexity in de Sitter Space},
  Class.\ Quant.\ Grav.\  {\bf 34}, no. 17, 175013 (2017)
  [arXiv:1706.03788 [hep-th]]

\bibitem{Nguyen:2017ggc} 
  K.~Nguyen,
{\it De Sitter-invariant States from Holography},
  Class.\ Quant.\ Grav.\  {\bf 35}, no. 22, 225006 (2018)
  [arXiv:1710.04675 [hep-th]].
  



\bibitem{Mahajan:2014daa} 
N.~Mahajan,
{\it BICEP2, non-Bunch-Davies and entanglement},
Phys.\ Lett.\ B {\bf 743}, 403 (2015)
[arXiv:1405.3247 [hep-ph]]

\bibitem{Maldacena:2015bha} 
  J.~Maldacena,
 {\it A model with cosmological Bell inequalities},
  Fortsch.\ Phys.\  {\bf 64}, 10 (2016)
  [arXiv:1508.01082 [hep-th]].

\bibitem{Boyanovsky:2018soy} 
  D.~Boyanovsky,
  {\it Imprint of entanglement entropy in the power spectrum of inflationary fluctuations},
  Phys.\ Rev.\ D {\bf 98}, no. 2, 023515 (2018)
[arXiv:1804.07967 [astro-ph.CO]]



\bibitem{Choudhury:2018fpj} 
S.~Choudhury and S.~Panda,
{\it Spectrum of cosmological correlation from vacuum fluctuation of Stringy Axion in entangled de Sitter space},
arXiv:1809.02905 [hep-th]

  
  \bibitem{Kanno:2018cuk} 
  S.~Kanno and J.~Soda,
{\it Possible detection of nonclassical primordial gravitational waves with Hanbury Brown - Twiss interferometry},
  arXiv:1810.07604 [hep-th]
  

  
  
  
  
  \bibitem{Kanno:2016qcc} 
    S.~Kanno, M.~Sasaki and T.~Tanaka,
 {\it Vacuum State of the Dirac Field in de Sitter Space and Entanglement Entropy},
  JHEP {\bf 1703}, 068 (2017)
  [arXiv:1612.08954 [hep-th]]
  

  
    
  \bibitem{Fuentes:2010dt} 
  I.~Fuentes, R.~B.~Mann, E.~Martin-Martinez and S.~Moradi,
  {\it Entanglement of Dirac fields in an expanding spacetime},
  Phys.\ Rev.\ D {\bf 82}, 045030 (2010)
  [arXiv:1007.1569 [quant-ph]]
  
  \bibitem{Kwon:2015gaa} 
  Y.~Kwon,
{\it No survival of Nonlocalilty of fermionic quantum states with alpha vacuum in the infinite acceleration limit},
Phys.\ Lett.\ B {\bf 748}, 204 (2015)

\bibitem{Machado:2018lyn} 
  L.~N.~Machado, H.~A.~S.~Costa, I.~G.~Da Paz, M.~Sampaio and J.~B.~Araujo,
 {\it Interacting fermions in an expanding spacetime},
  arXiv:1811.00575 [hep-th].

\bibitem{Alsing:2006cj} 
  P.~M.~Alsing, I.~Fuentes-Schuller, R.~B.~Mann and T.~E.~Tessier,
  {\it Entanglement of Dirac fields in non-inertial frames},
  Phys.\ Rev.\ A {\bf 74}, 032326 (2006)
  [quant-ph/0603269]
  
  
  \bibitem{Montero:2011ai} 
  M.~Montero and E.~Martin-Martinez,
  {\it Fermionic entanglement ambiguity in non-inertial frames},
  Phys.\ Rev.\ A {\bf 83}, 062323 (2011)
  [arXiv:1104.2307 [quant-ph]]

  
  \bibitem{Wald}
R.~M.~Wald, {\it General Relativity}, Chicago Univ. Press (1984)

  
\bibitem{Blasone}
M.~Blasone, {\it Canonical Transformations in Quantum Field Theory}, Lecture notes :       
  http://www.sa.infn.it/massimo.blasone/


  
  
    
  
  
  





   \bibitem{Gromes:1974yu} 
  D.~Gromes, H.~J.~Rothe and B.~Stech,
{\it Field quantization on the surface $\chi$-squared = constant},
  Nucl.\ Phys.\ B {\bf 75}, 313 (1974)
  
  \bibitem{Camporesi:1995fb} 
  R.~Camporesi and A.~Higuchi,
  {\it On the Eigen functions of the Dirac operator on spheres and real hyperbolic spaces},
  J.\ Geom.\ Phys.\  {\bf 20}, 1 (1996)
  [gr-qc/9505009]

  
   
  




\bibitem{Copson}
E.~T.~Copson, {\it An Introduction to the Theory of Functions of a Complex Variable}, Oxford, Clarendon
(1960)


\bibitem{Yamauchi:2014saa} 
  D.~Yamauchi, T.~Fujita and S.~Mukohyama,
  {\it Is there supercurvature mode of massive vector field in open inflation?},
  JCAP {\bf 1403}, 031 (2014)
  [arXiv:1402.2784 [astro-ph.CO]]



\bibitem{Ashoorioon:2014nta} 
  A.~Ashoorioon, K.~Dimopoulos, M.~M.~Sheikh-Jabbari and G.~Shiu,
{\it Non-Bunch-Davis initial state reconciles chaotic models with BICEP and Planck},
  Phys.\ Lett.\ B {\bf 737}, 98 (2014)
 [arXiv:1403.6099 [hep-th]]
 


  


  
  
  \bibitem{Bytsenko:1994bc} 
  A.~A.~Bytsenko, G.~Cognola, L.~Vanzo and S.~Zerbini,
  Phys.\ Rept.\  {\bf 266}, 1 (1996)
  [hep-th/9505061]
  
  
  
   
  
  
  
   
    

   
  
  \end{thebibliography}
\end{document}